\documentclass[aps,preprint,showpacs,showkeys,preprintnumbers,amsmath,amssymb]{revtex4}

\usepackage{dcolumn}
\usepackage{mathrsfs}
\usepackage{bm}
\usepackage{amsmath,amssymb,epsfig,float}

\begin{document}

\title{How the Unruh effect affects  transition between
classical and quantum decoherences}

\author{Zehua Tian and Jiliang Jing\footnote{Corresponding  author, Email: jljing@hunnu.edu.cn}}
\affiliation{ Department of Physics, and
 Key Laboratory of Low-dimensional Quantum Structures
\\ and Quantum
Control of Ministry of Education, \\ Hunan Normal University,
Changsha, Hunan 410081, P. R. China}

\vspace*{0.2cm}
\begin{abstract}
\vspace*{0.2cm}

We investigate how the Unruh effect affects the transition  between
classical and quantum decoherences for a general class of initial
states and find that: $(i)$  The quantum decoherence exists while
$\lambda t\leq\lambda \widetilde{t}$ (the transition time) and the
classical one can also affect the system's evolution while $\lambda
t\geq\lambda\widetilde{t}$ for both the bit and phase-bit flips,
which are different from the cases in inertial frame; $(ii)$ The classical decoherence will not occur, while the quantum decoherence still dominates the evolution of system as  $\lambda t\geq\lambda\widetilde{t}$ for the phase flip;
And $(iii)$ as the Unruh temperature increases, the $\lambda\widetilde{t}$, compared
with that in inertial frame, will be bigger for phase flip but smaller for bit flip. However, the $\lambda\widetilde{t}$ does not change no matter what the Unruh effect is for phase-bit flip.
\end{abstract}

\vspace*{0.5cm}


\maketitle

\section{introduction}
The relativistic quantum information plays a  key role in better
understanding the quantum mechanics \cite{Peres,Bouwmeester} and is
helpful to study the information paradox existing in black hole
\cite{Bombelli-Callen,Hawking-Terashima}. Because of that, it has
attracted much attention recently \cite{P. M. Alsing,Fuentes,R. B.
Mann,Q. Pan,J. Deng,L. J. Garay,J. Leon,Wang Jing,Montero,J. Louko}.
However, because the interaction between the quantum system and
surrounding environment is inevitable, the study of the interaction
becomes an important topic in quantum information, and more and more
authors focus their attention on it \cite{Zurek,Breuer}. Especially,
J. Maziero \emph{et al.} \cite{Maziero} discussed classical and
quantum correlations under decoherence and identified three types of
dynamics that include a peculiar sudden change in their decay rates.
And they showed that, under suitable conditions, the classical
correlation is unaffected by decoherence. L. Mazzola \emph{et al.}
\cite{Mazzola} pointed out that there exists a sudden transition
between the classical and quantum decoherences for a class of
initial states when system experiences a phase flip course, which is the first evidence of the existence of quantum properties. Here, the classical decoherence, which was introduced by L. Mazzola etc. in Ref. \cite{Mazzola}, represents the decoherence process inducing loss of classical correlation. Most recently J. Wang \emph{et al.} \cite{Jieci} studied quantum
decoherence in noninertial frames, and they found that the sudden
death of entanglement, in the case of the total system under
decoherence, may appear for any acceleration for Dirac field.
However, in these papers the authors either considered the quantum
and classical decoherences only in inertial frame or do not
distinguish the quantum and classical decoherences in noninertial
frames. Here we will extend the study of \cite{Mazzola} to the
noninertial frames and will present some interesting new conclusions
of the quantum and classical decoherences due to the Unruh effect.

The structure of the paper is as follows. In Sec. II we recall some
concept of the classical and quantum correlations. In Sec. III we
introduce the essential features of the Dirac fields in the
noninertial frame simply. In Sec. IV we study the transition between
classical and quantum correlations in noninertial frames. And we
summarize and discuss our conclusions in the last section.

\section{Classical and quantum correlations}

We now introduce quantum discord and classical correlation briefly.
As we all known, Shannon entropy is always used to quantify
information in classical information theory, which is defined as
$H(X)=-\sum_x P_{|X=x} \log P_{|X=x}$, where $P_{|X=x}$ is the
probability with $X$ being $x$. And mutual information, to quantify
the relationship between two random variables $X$ and $Y$ , is
introduced here, whose formula is $I(X:Y)=H(X)+H(Y)-H(X,Y)$. Since
$H(Y|X)=H(Y,X)-H(X)$, there is another alternative expression for
the mutual information $I(X:Y)=H(Y)-H(Y|X)$, where $H(Y|X)$ is
conditional entropy. Classically, these two expression of mutual
information are identical. Now we generalize these expressions to
quantum domain. Their quantum version can be written as \cite{RAM}
\begin{equation}
{\cal I}(A:B)=S(\rho_A)+S(\rho_B)-S(\rho_{AB}) \label{mi2},
\end{equation}
and
\begin{equation}
{\cal J}_{\{\Pi_j\}}(A:B)=S(\rho_B)-S_{\{\Pi_j\}}(B|A) \label{mi3},
\end{equation}
where $S(\rho)=-{\rm Tr}(\rho {\rm log}\rho)$ is the von Neumann
entropy, $S_{\{\Pi_j\}}(B|A)=\sum_j p_j S(\rho_{B|j})$ is conditional entropy \cite{Cerf} of quantum state, and $\{\Pi_j\}$ is a complete set of projectors, which is used to measure subsystem $A$ of $\rho_{AB}$. And the outcome is $j$, corresponding to $\rho_{B|j}={Tr_A(\Pi_j\rho_{AB}\Pi_j)}/{p_j}$, with ${p_j}=Tr_{AB}(\Pi_j\rho_{AB}\Pi_j)$.  Incidentally, ${\cal I}(A:B)$ and ${\cal J}(A:B)$ do not equal each other in this case.

From above we know that conditional entropy is strongly effected by the choice of the measurements $\{\Pi_j\}$. Hence, we, in order to get the classical correlation, have to minimize the conditional entropy over all possible measurements of $A$, and it needs to find a optimal measurement which disturb least the overall system \cite{Ollivier,Shunlong,Shun}. Then we define the classical correlation between two subsystems $A$ and $B$ as
\begin{equation}
{\cal C}(A:B)=\max_{\{\Pi_j\}}{\cal J}_{\{\Pi_j\}}(A:B),
\end{equation}
and quantum discord \cite{Ollivier,Shun}

\begin{equation}
{\cal D}(A:B)={\cal I}(A:B)-{\cal C}(A:B).
\end{equation}

 Quantum discord is nonnegative and is zero for states with only classical correlations \cite{Zurek,Ollivier}. Thus a nonzero value of ${\cal D}(A:B)$ indicates the presence of nonclassical correlation \cite{Wehrl,Stratonovich}, that is, there exists quantum correlation.

\section{Quantization of Dirac filed in Minkowski and
Rindler spacetimes}

 For a inertial observer in flat Minkowski spacetime, we can expand the free field in terms of the positive (particle) and the negative (antiparticle) energy solution of Dirac equation
\begin{equation}
\Psi=\sum_i\int d k \, ( a_{k_i}\, \psi^+_{k_i} + b^+_{k_i}\,\psi^-_{k_i} ),
\end{equation}
where the subscript $k$ is momentum notation,  which is used to
label the modes with the same energy, with $\omega_i=| k_i|$
corresponding to massless Dirac fields. The particle and
antiparticle operators satisfy the usual anticommutation rule
\begin{equation}
\{a_{k_i},a^+_{k_j}\}=\{b_{k_i},b^+_{k_j}\}=\delta(k_i-k_j),
\end{equation}
and all other anticommutators vanishing.

For a accelerating observer, Rindler coordinates $(\eta, \varepsilon)$ are the appropriate coordinates to describe him. The relationship between the Minkowski coordinates and Rindler coordinates is
\begin{eqnarray}
at=e^{a \varepsilon}\ \sinh {(a\eta)}, \ \ \ \    az=e^{a \varepsilon}\ \cosh {(a\eta)},
\\ \nonumber
at=-e^{a \varepsilon}\ \sinh {(a\eta)}, \ \ \ \    az=-e^{a \varepsilon}\ \cosh {(a\eta)}
\end{eqnarray}
for region $\mathrm{I}$ and $\mathrm{II}$, respectively.

We now expand the Dirac field in the terms of Rindler modes, which is given by
\begin{equation}
\Psi=\sum_i\int
d k [\hat{c}^I_{k_i}\Psi^{I+}_{k_i}+\hat{d}^{I+}_{k_i}\Psi^{I-}_{k_i}+
\hat{c}^{II}_{k_i}\Psi^{II+}_{k_i}+\hat{d}^{II+}_{k_i}\Psi^{II-}_{k_i}],
\end{equation}
where $\hat{c}^{s}_{k_i}$ and $\hat{d}^{s+}_{k_i}$ are the fermion annihilation and antifermion creation operators of state in region $s$ respectively, with s=${\{I,II}\}$. The anticommutation relations of these operators are
\begin{equation}
\{\ \hat{c}^{s}_{k_i},\hat{c}^{s'+}_{k_j}\}=
\{\ \hat{d}^{s}_{k_i},\hat{d}^{s'+}_{k_j}\}=
\delta(k_i-k_j)\delta_{ss'}.
\end{equation}

According to the Bogoliubov transformation,  in the Rindler
coordinates we can easily get Minkowski vacuum state
$|0\rangle_{M}=\bigotimes_i|0_{\omega_i}\rangle_{M} $ and excited
state $|1\rangle_{M}=\bigotimes_i|1_{\omega_i}\rangle_{M} ~\forall
i$, with
\begin{eqnarray}\label{Dirac-vacuum}
&&|0_{w_i}\rangle_{M}=(e^{-\omega_i/T}+1)^{-\frac{1}{2}}|0_{w_i}\rangle_{I}|0_{w_i}\rangle_{II}
+(e^{\omega_i/T}+1)^{-\frac{1}{2}}|1_{w_i}\rangle_{I}|1_{w_i}\rangle_{II}, \\
&&|1_{w_i}\rangle_{M}=|1_{w_i}\rangle_{I}|0_{w_i}\rangle_{II},
\end{eqnarray}
where $T=a/2\pi $ is the Unruh temperature.

From Eq. (10), we can see that the Minkowski vacuum, from the perspective of the uniformly accelerated observer Bob, is a two-mode squeezed state of the Rindler Fock state. Usually, we think that Bob and antiBob are confined in region $\mathrm{I}$ and $\mathrm{II}$, respectively, which are causally disconnected. Therefore, the observer must trace over the inaccessible region, as a result of that, the information will be lost, which essentially results in the detection of a thermal state. That is so called Unruh effect. Because of the Unruh effect, a entangled pure state seen by the inertial observer Alice appears mixed from the accelerated frame.

\section{transition between
classical  and quantum decoherences}

We consider a maximally mixed marginal  initial state shared by Alice and Bob, which can be
expressed as
\begin{equation}
\rho_{AB}=\frac{1}{4}\left({\mathbf1}_{AB} +\sum_{i=1}^3
c_i\sigma^A_i\otimes\sigma^B_i\right),
\end{equation}
where $\sigma^n_i$ is the standard Pauli  operator in direction $i$
acting on the subspace $n={\{A,B}\}$, $c_i\in\Re$ with $0\leq\mid
c_i\mid\leq1$ for $i={1,2,3}$ and ${\mathbf1}_{A(B)}$ is the
identity operator in subspace $A(B)$. It is obvious that different
coefficients represent different sates. Especially,
$|c_1|=|c_2|=|c_3|=1$ and $|c_1|=|c_2|=|c_3|=c$ are corresponding to Bell basic state and Werner state, respectively.

Now we assume that one observer Bob moves with an uniform acceleration while Alice stays stationary. At the same time, Alice has a detector only sensitive to mode$\mid n\rangle_A $ and Bob has a detector only sensitive to mode $\mid n\rangle_B$, respectively. Then we use Eqs. $(10)$ and $(11)$ to rewrite the Eq. $(13)$ in terms of Minkowski modes for Alice and Rindler modes for Bob and trace over the state in region $II$, we get
\begin{eqnarray}
\rho_{A,I}=\left(
  \begin{array}{cccc}
    \frac{(1+c_3)}{4(e^{-\omega/T}+1)} & 0 & 0 & \frac{(c_1-c_2)}{4(e^{-\omega/T}+1)^{\frac{1}{2}}} \\
    0 & \frac{(1-c_3)+(1+c_3)(e^{\omega/T}+1)^{-1}}{4} & \frac{(c_1+c_2)}{4(e^{-\omega/T}+1)^{\frac{1}{2}}} & 0 \\
    0 & \frac{(c_1+c_2)}{4(e^{-\omega/T}+1)^{\frac{1}{2}}} & \frac{(1-c_3)}{4(e^{-\omega/T}+1)} & 0 \\
    \frac{(c_1-c_2)}{4(e^{-\omega/T}+1)^{\frac{1}{2}}} & 0 & 0 & \frac{(1+c_3)+(1-c_3)(e^{\omega/T}+1)^{-1}}{4}
  \end{array}
\right).
\end{eqnarray}

Now we consider the case of two qubits under local decoherence channels. If the initial state $\rho_{AB}$ undergoes decoherence environment, then the evolved state can be represented as the operator-sum version \cite{Maziero}
\begin{equation}
\varepsilon(\rho_{AB})=\sum_{i,j}\Gamma^{(A)}_i\Gamma^{(B)}_j \rho_{AB}\Gamma^{(B)\dag}_i\Gamma^{(A)\dag}_j,
\end{equation}
where $\Gamma^{(k)}_i(k=A,B)$ are the Kraus operators that describe the noise channels for subsystems $A$ and $B$.

Ref. \cite{Mazzola,Maziero} found that there exists a sudden
transition between the classical and quantum decoherences for phase
flip when $|c_1|\geq|c_2|, |c_3|$ or $|c_2|\geq|c_1|, |c_3|$ and
$|c_3|\neq0$, for bit flip when $|c_3|\geq|c_1|, |c_2|$ or
$|c_2|\geq|c_1|, |c_3|$ and $|c_1|\neq 0$, and for phase-bit flip
when $|c_3|\geq|c_1|, |c_2|$ or $|c_1|\geq|c_3|, |c_2|$ and
$|c_2|\neq0$. Under these conditions, by analyzing  quantum discord
and  classical correlation of the three noise channels in the
noninertial frame, we can further understand the decoherence.

\subsection{Phase flip}

The phase flip channel is a quantum noise process with loss of quantum information without loss of energy \cite{Maziero}. Its Kraus operators are given by \cite{Nielsen} $\Gamma^{(A)}_0=diag(\sqrt{1-p_A/2},\sqrt{1-p_A/2})\otimes\mathbf{1_B}$, $\Gamma^{(A)}_1=diag(\sqrt{p_A/2},-\sqrt{p_A/2})\otimes\mathbf{1_B}$, $\Gamma^{(B)}_0=\mathbf{1_A}\otimes diag(\sqrt{1-p_B/2},\sqrt{1-p_B/2})$, $\Gamma^{(B)}_1=\mathbf{1_A}\otimes diag(\sqrt{p_B/2},-\sqrt{p_B/2})$,
where $p_{A(B)}(0\leq p_{A(B)} \leq 1)$ is function of time. For simplicity, we assume $p_A=p_B=p$ throughout this paper. Then we have
\begin{eqnarray}
\nonumber\\\varepsilon(\rho_{A,I})&=&\frac{1}{4}\left(\mathbf{1}_{A,I}+c^{\prime}_0 \mathbf{1}_{A}\otimes\sigma^\mathbf{I}_3+\sum_{i=1}^3 c^{\prime}_i\sigma^A_i\otimes\sigma^\mathbf{I}_i\right),\label{eab}
\end{eqnarray}
where $c^{\prime}_0=\frac{-1}{(e^{\omega/T}+1)}$, $c^{\prime}_1=\frac{(1-p)^2c_1}{(e^{-\omega/T}+1)^\frac{1}{2}}$, $c^{\prime}_2=\frac{(1-p)^2c_2}{(e^{-\omega/T}+1)^\frac{1}{2}}$ and  $c^{\prime}_3=\frac{c_3}{(e^{-\omega/T}+1)}$. Because of $p=1-\exp(-\lambda t)$, where $\lambda$ is the phase damping rate \cite{Maziero}, we can rewrite $c^\prime_1$ and $c^\prime_2$ as  $c^\prime_1(t)=\frac{c_1(0)\exp(-2\lambda t)}{(e^{-\omega/T}+1)^\frac{1}{2}}$ and  $c^\prime_2(t)=\frac{c_2(0)\exp(-2\lambda t)}{(e^{-\omega/T}+1)^\frac{1}{2}}$.

For the sake of getting the quantum and classical correlation, we first need to calculate the eigenvalues of the state $\varepsilon(\rho_{A,I})$. From Eq. (\ref{eab}) we have
\begin{eqnarray}
\begin{array}{cccc}
\lambda_1=\frac{1}{8}[2-c_3-c_3(\frac{1-e^{-\omega/T}}{1+e^{-\omega/T}})-2\sqrt{\frac{(c_1+c_2)^2e^{-4\lambda t}}{e^{-\omega/T}+1}+\frac{1}{(e^{\omega/T}+1)^2}}],
\\ \lambda_2=\frac{1}{8}[2-c_3-c_3(\frac{1-e^{-\omega/T}}{1+e^{-\omega/T}})+2\sqrt{\frac{(c_1+c_2)^2e^{-4\lambda t}}{e^{-\omega/T}+1}+\frac{1}{(e^{\omega/T}+1)^2}}],
\\ \lambda_3=\frac{1}{8}[2+c_3+c_3(\frac{1-e^{-\omega/T}}{1+e^{-\omega/T}})-2\sqrt{\frac{(c_1-c_2)^2e^{-4\lambda t}}{e^{-\omega/T}+1}+\frac{1}{(e^{\omega/T}+1)^2}}],
\\ \lambda_4=\frac{1}{8}[2+c_3+c_3(\frac{1-e^{-\omega/T}}{1+e^{-\omega/T}})+2\sqrt{\frac{(c_1-c_2)^2e^{-4\lambda t}}{e^{-\omega/T}+1}+\frac{1}{(e^{\omega/T}+1)^2}}].
\end{array}
\end{eqnarray}
And we can also get the entropy $s(\rho_A)$ for the reduced density matrix of mode $A$ and $s(\rho_I)$ for the mode $I$, respectively. Then the mutual information between $A$ and $I$ is
\begin{eqnarray}
{\cal I}(\rho_{A,I})&=&S(\rho_A)+S(\rho_I)-S(\rho_{A,I})
\nonumber \\
&=&1-\frac{1}{2(e^{-\omega/T}+1)}\log_2(\frac{1}{2(e^{-\omega/T}+1)})-\frac{1+(e^{\omega/T}+1)^{-1}}{2} \log_2(\frac{1+(e^{\omega/T}+1)^{-1}}{2})
\nonumber \\
&+&\sum_{i=1}^4\lambda_i\log_2(\lambda_i).\label{mi22}
\end{eqnarray}

Now we calculate the conditional entropy, which is the key of calculating the classical and quantum correlation. Let us make our  measurements on the subsystem $A$ first. The projectors are defined as \cite{Jun}
\begin{eqnarray}
\Pi_{+}=\frac{I_1+\mathbf{n}\cdot\sigma}{2}\otimes I_2,\ \ \ \ \
\Pi_{-}=\frac{I_1-\mathbf{n}\cdot\sigma}{2}\otimes I_2,
\end{eqnarray}
where, in spherical coordinates, $n_1=\sin\theta\cos\varphi$, $n_2=\sin\theta\sin\varphi$, $n_3=\cos\theta$, and $\sigma_i$ are the Pauli matrices. After the measurements, the final states are
\begin{eqnarray}\label{con1}
&&\rho_{(I|+)}=Tr_{A}(\Pi_{+}\rho_{A,I}\Pi_{+})/p_{+} \nonumber\\
&&=\frac{1}{2}\left[
\begin{array}{cc}
\frac{(1+c_3\cos\theta)}{e^{-\omega/T}+1} & \frac{(c_1\cos\varphi-ic_2\sin\varphi)e^{-2\lambda t }\sin\theta}{(e^{-\omega/T}+1)^{\frac{1}{2}}}\\
 \frac{(c_1\cos\varphi+ic_2\sin\varphi)e^{-2\lambda t }\sin\theta}{(e^{-\omega/T}+1)^{\frac{1}{2}}}& 2-\frac{1+c_3\cos\theta}{e^{-\omega/T}+1} \\
\end{array}
\right],
\end{eqnarray}
and
\begin{eqnarray}\label{con12}
&&\rho_{(I|-)}=Tr_{A}(\Pi_{-}\rho_{A,I}\Pi_{-})/p_{-} \nonumber \\
&&=\frac{1}{2}\left[
\begin{array}{cc}
\frac{(1-c_3\cos\theta)}{e^{-\omega/T}+1} & \frac{-(c_1\cos\varphi-ic_2\sin\varphi)e^{-2\lambda t }\sin\theta}{(e^{-\omega/T}+1)^{\frac{1}{2}}}\\
 \frac{-(c_1\cos\varphi+ic_2\sin\varphi)e^{-2\lambda t }\sin\theta}{(e^{-\omega/T}+1)^{\frac{1}{2}}}& 2-\frac{1-c_3\cos\theta}{e^{-\omega/T}+1} \\
\end{array}
\right],
\end{eqnarray}
where $p_{+}=Tr(\Pi_{+}\rho_{A,I}\Pi_{+})=1/2$ , $p_{-}=Tr(\Pi_{-}\rho_{A,I}\Pi_{-})=1/2$. Then the eigenvalues of state $\rho_{(I|+)}$ and $\rho_{(I|-)}$ are given by
\begin{eqnarray}
\lambda_{+}(1,2)=\frac{1}{2}(1\pm\sqrt{(\frac{1+c_3\cos \theta}{e^{-\omega/T}+1}-1)^2+\frac{\sin^2\theta(c^2_1\cos^2\varphi+c^2_2\sin^2\varphi)e^{-4\lambda t}}{e^{-\omega/T}+1}} ),
\end{eqnarray}
and
\begin{eqnarray}
\lambda_{-}(1,2)=\frac{1}{2}(1\pm\sqrt{(\frac{c_3\cos \theta-1}{e^{-\omega/T}+1}+1)^2+\frac{\sin^2\theta(c^2_1\cos^2\varphi+c^2_2\sin^2\varphi)e^{-4\lambda t}}{e^{-\omega/T}+1}} ).
\end{eqnarray}
Now it is easily to obtain the conditional entropy $S_{\{\Pi_j\}}(I|A)=\sum_jp_jS(I|j)$. Minimizing the conditional entropy, we can get the classical correlation
\begin{eqnarray}
{\cal C}(\rho_{A,I})=&-&\frac{1}{2(e^{-\omega/T}+1)}\log_2(\frac{1}{2(e^{-\omega/T}+1)})
-\frac{1+(e^{\omega/T}+1)^{-1}}{2} \log_2(\frac{1+(e^{\omega/T}+1)^{-1}}{2})
\nonumber \\
&-&\min_{\Pi_j}S_{\{\Pi_j\}}(I|A),
\end{eqnarray}
and quantum discord
\begin{eqnarray}{\cal D}(\rho_{A,I})=1+\sum_{i=1}^4\lambda_i\log_2(\lambda_i)+\min_{\Pi_j}S_{\{\Pi_j\}}(I|A).
\end{eqnarray}

Obviously, the conditional entropy has to be numerically evaluated by optimizing over the angles $\theta$ and $\varphi$. To minimize it, we need to consider all the measurements on $A$ and find the optimizing measurement. According to the analysis, we can get two values of $\varphi$ for optimizing measurement $\varphi=0$ and $\varphi=\frac{\pi}{2}$ which corresponds to $|c_1|\geq|c_2|$ and $|c_2|\geq|c_1|$ \cite{Shun}, respectively. We now plot condition entropy as a function of $\theta$ and $\lambda t$ $($to compare with \cite{Mazzola,Maziero}, we also take the values of $c_1=1$, and $-c_2=c_3=0.6$ $)$ in Fig.1.
\begin{figure}[ht]
\includegraphics[scale=0.75]{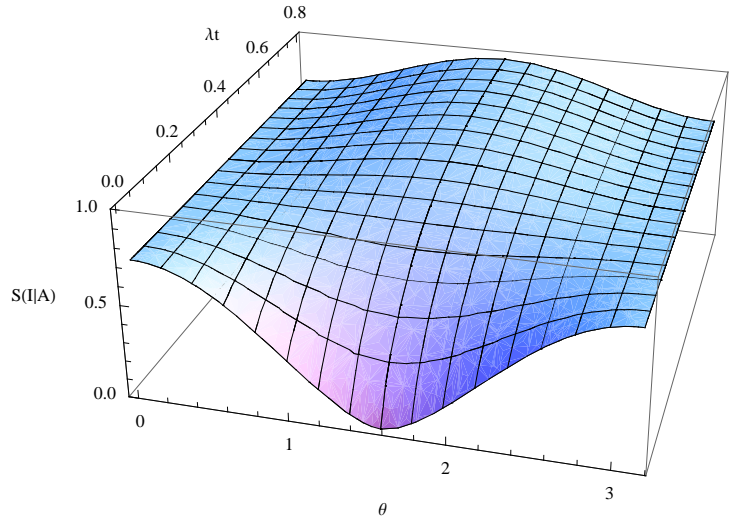}\vspace{0.0cm}
\caption{\label{CE}(Color online) The conditional entropy $S(I|A)$
as a functions of $\lambda t$ and $\theta$ (the diagram for  $T=0$).}
\end{figure}
From which we know that the condition of obtaining the minimum conditional entropy is $\theta=\frac{\pi}{2}$ for $0\leq\lambda t\leq\lambda \widetilde{t}$ and $\theta=0$ for $\lambda \widetilde{t}\leq\lambda t\leq 1$, where the turning point is $\lambda \widetilde{t}_{(T=0)}=0.25541$. We can also get the approximate turning points $\lambda \widetilde{t}_{(T=\frac{1}{2\ln(\cot\frac{\pi}{8})})}=0.29024$, $\lambda \widetilde{t}_{(T=\frac{1}{2\ln(\cot\frac{\pi}{6})})}=0.30387$ and $\lambda \widetilde{t}_{(T\rightarrow\infty)}=0.37326$ (for simplicity, we assume $w=1$).

We present the mutual information for differen Unruh temperature by Fig.$2$, which shows the evolution of the mutual information as time increases.
Obviously, the mutual information will decrease as time increases, and the values are smaller for the higher Unruh temperature with the same time. However, because the mutual information includes both the classical and quantum correlation,  we can not get the classical property or quantum property from the mutual information directly. That is, we can not read out the quantum decoherence or classical one, which are responsible for the decrease of the information,  from it.
\begin{figure}[ht]
\includegraphics[scale=0.75]{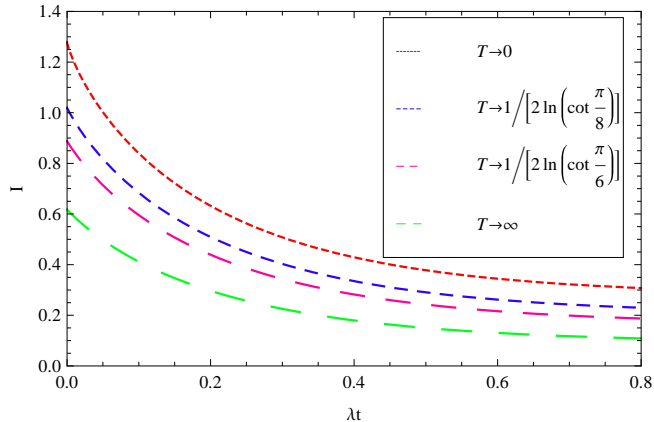}\vspace{0.0cm}
\caption{\label{MI}(Color online) The mutual information $\cal I$ as a function of $\lambda t$ for different Unruh temperature with $c_1=1$, $-c_2=c_3=0.6$.}
\end{figure}

To understand the quantum and classical decoherences better, in Fig. $3$ we plot the time evolution of quantum discord and the classical correlation for differen Unruh temperatures.
\begin{figure}[ht]
\includegraphics[scale=0.75]{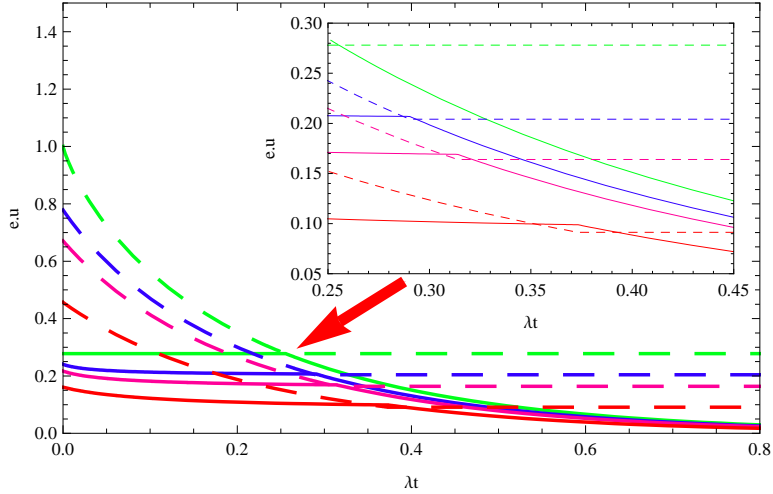}\vspace{0.0cm}
\caption{\label{CD}(Color online) The classical correlation $\cal C$ (dashed line) and quantum discord $\cal D$ (solid line) as  functions of $\lambda t$ for different Unruh effect (from the top to bottom are corresponding to $T=0$, $T=\frac{1}{2\ln(\cot\frac{\pi}{8})}$, $T=\frac{1}{2\ln(\cot\frac{\pi}{6})}$, $T\rightarrow\infty$,) with $c_1=1$, $-c_2=c_3=0.6$. In the inset we plot the detail of each transition.}
\end{figure}
For $T=0$, the time point $\lambda \widetilde{t}_{(T=0)}=0.25541$ is a interesting point, where there is a sharp transition from the classical to the quantum decoherences \cite{Maziero,Mazzola}. When $\lambda t_{(T=0)}\leq\lambda\widetilde{t}_{(T=0)}$, only the classical correlation decreases, while the quantum discord is constant in time, so the decoherence process is the loss of only classical correlation. Obviously, for $\lambda\widetilde{t}_{(T=0)}\leq\lambda t_{(T=0)}$, classical correlation does not change in time while only quantum decoherence occurs.

Consider that the classical and quantum decoherences regimes are strictly distinguished in this case, it is very useful to study how the Unruh effect affect the classical and quantum decoherences.  From the Fig. $3$ we note that both the classical correlation and quantum discord are piecewise functions, and $\lambda \widetilde{t}$ is the breakpoint. The figure also shows that: $(i)$ Because of the existence of Unruh effect, the quantum decoherence will happen in the time interval $\lambda t_{(T\neq0)}\leq\lambda\widetilde{t}_{(T\neq0)}$, which is unlike the case of $T=0$. $(ii)$ It is very interesting to note that the classical decoherence doesn't occur when $\lambda t_{(T\neq0)}\geq\lambda \widetilde{t}_{(T\neq0)}$, which is the same with the case of inertial fame. $(iii)$ We also find that the transition between classical and quantum decoherences will delay with the Unruh temperature increases. And $(iv)$  from the inset, we know that the quantum discord will become larger than the classical correlation near the $\lambda\widetilde{t}$, the higher the Unruh temperature is, the longer the time interval is.

For future understanding the dynamics of the total quantum correlation, we will calculate the concurrence $\cal CC$ which is defined as ${\cal CC}=\max\{0,\sqrt{\lambda_1}-\sqrt{\lambda_2}-\sqrt{\lambda_3}-\sqrt{\lambda_4}\}, \lambda_i\geq\lambda_{i+1}\geq 0$ \cite{Wootters,Coffman}, where $\sqrt{\lambda_i}$ are square root of the eigenvalues of the matrix $\rho\widetilde{\rho}$ with $\widetilde{\rho}=(\sigma_y\otimes\sigma_y)\rho^*(\sigma_y\otimes\sigma_y)$ defined as $``$ spin-flip$"$ matrix of the state $\rho$.
\begin{figure}[ht]
\includegraphics[scale=0.75]{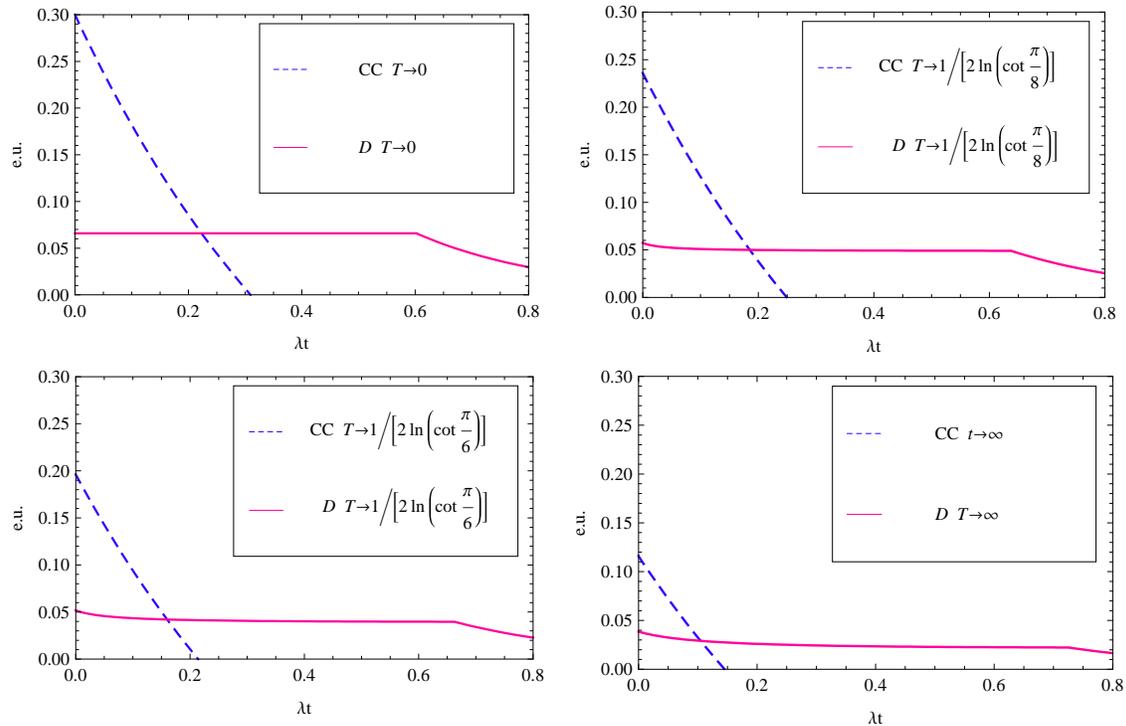}\vspace{0.0cm}
\caption{\label{DCC}(Color online) The concurrence $\cal CC$ (dashed line) and quantum discord $\cal D$ (solid line)
as functions of $\lambda t$ with $c_1=1$, $-c_2=c_3=0.3$.}
\end{figure}

Fig.4 is a compare of concurrence $\cal CC$ and discord $\cal D$. We find entanglement decreases monotonically. When $\lambda t\geq\lambda t_S= -\frac{1}{2}\ln\frac{\sqrt{(-1+c_3)[c_3- 3+(1+c_3)(\frac{1-e^{-\omega/T}}{1+e^{-\omega/T}})]}}{\sqrt{2}(c_1-c_2)}$ $($ this solution for the case of $c_1>c_3, -c_2>0$, or exchanging the role of $c_1$ with $c_2$ $)$, the entanglement will disappears completely, but quantum discord does not. For inertial frame $($T=0$)$, if $\lambda t_S\leq\lambda\widetilde{t}$, entanglement vanishes when the quantum discord has not yet started to decay so the state of the total system is separable state with nonzero-discord \cite{Maziero,Mazzola}. It is interesting note that  this conclusion is also applicable to the noninertial frame, although its quantum discord will weakly decays when $\lambda t\leq\lambda\widetilde{t}$.  The time interval $\lambda\widetilde{t}-\lambda t_S$ increases as the Unruh temperature increases, that is, the Unruh effect can induce that sudden death of the entanglement happens earlier and  the transition between classical and quantum decoherences occurs later. Ref. \cite{Mazzola} pointed that the entanglement vanishes while the quantum discord has not yet started to decay only for the condition of $0<|c_3|<\sqrt{2}-1$. Here, we find that the upper bound for $\lambda t_S\leq\lambda\widetilde{t}$ becomes larger due to the Unruh effect.

\subsection{Bit flip}

For bit flip channel,  Ref. \cite{Mazzola,Maziero} found that there exists the sudden transition from classical decoherence to quantum one in the inertial frame. How the Unruh effect affect the classical and quantum decoherences in this channel? Our study, in Fig. \ref{f5}, shows that:
\begin{figure}[ht]
\includegraphics[scale=0.75]{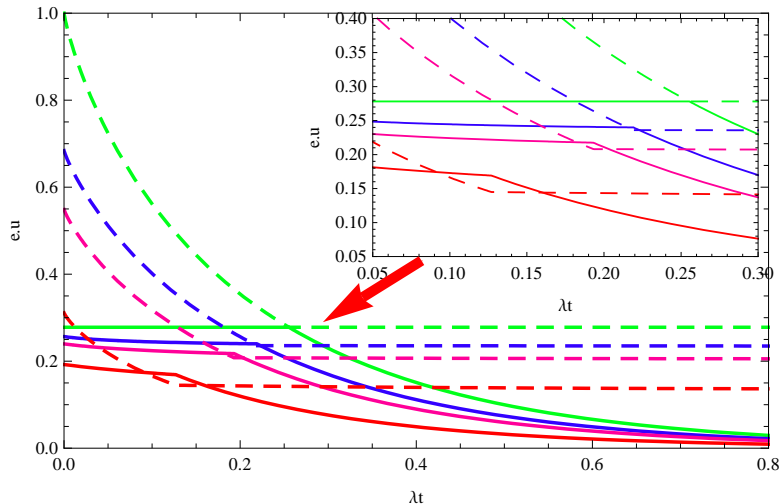}\vspace{0.0cm}
\caption{\label{BDC}(Color online) The classical correlation $\cal C$ (dashed line) and quantum discord $\cal D$ (solid line) as  functions of $\lambda t$ for  different Unruh effect (from the top to bottom are corresponding to $T=0$, $T=\frac{1}{2\ln(\cot\frac{\pi}{8})}$, $T=\frac{1}{2\ln(\cot\frac{\pi}{6})}$, $T\rightarrow\infty$,) with $c_3=1$, $-c_2=c_1=0.6$. In the inset we plot the detail of each transition.}\label{f5}
\end{figure}
$(i)$ It is very obvious that the transition from the classical to quantum decoherences occurs earlier with the Unruh temperature increases. $(ii)$ The classical decoherence also happens for time $\lambda t\geq\lambda\widetilde{t}$ with $T\neq 0$, but this phenomenon is too weak to be read from the graph directly. $(iii)$ The Unruh effect can also induce that the quantum discord becomes more than the classical correlation near the $\lambda\widetilde{t}$, the higher the Unruh temperature is, the longer the interval is. And $(iv)$ the Unruh effect can induce the quantum decoherence to happen before $\lambda\widetilde{t}$ too.

\subsection{Phase-bit flip}

\begin{figure}[ht]
\includegraphics[scale=0.75]{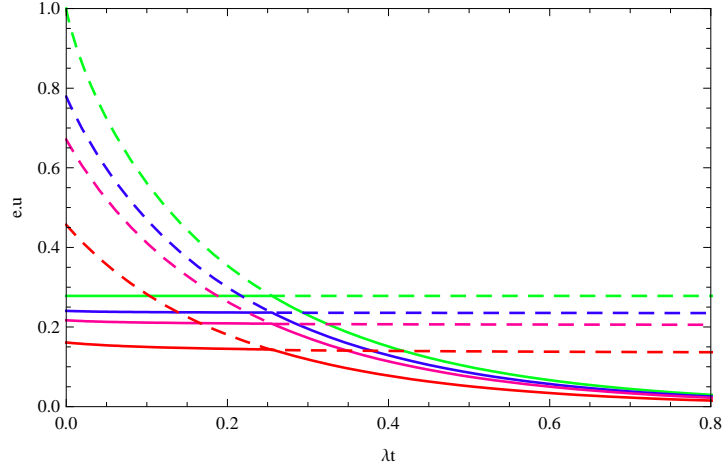}\vspace{0.0cm}
\caption{\label{PDC}(Color online) The classical correlation $\cal C$ (dashed line) and quantum discord $\cal D$ (solid line) as  functions of $\lambda t$ for different Unruh effect (from the top to bottom are corresponding to $T=0$, $T=\frac{1}{2\ln(\cot\frac{\pi}{8})}$, $T=\frac{1}{2\ln(\cot\frac{\pi}{6})}$, $T\rightarrow\infty$,) with $c_1=1$, $c_2=-c_3=0.6$.}\label{f6}
\end{figure}

In this case, the dynamic behavior of $\cal C$ and $\cal D$ under phase-bit flip in the noninertial frame is exhibited  in Fig. \ref{f6}.
And our study shows some interesting conclusions: $(i)$ Compared with the phase flip and bit flip, the biggest difference is that the transition time does not change no matter what the Unruh effect is, it is a constant, $\lambda\widetilde{t}=-\frac{1}{4}\ln\frac{c_2^2}{c_1^2}$. $(ii)$ The classical correlation is always larger than the quantum correlation except for $\lambda t=\lambda \widetilde{t}$. $(iii)$ The quantum decoherence can also exist before $\lambda\widetilde{t}$ as the consequence of the Unruh effect and the classical decoherence can exist after  $\lambda\widetilde{t}$ too.

\section{summary}
How the Unruh effect affects the transition  between the classical and quantum decoherences for a general class of initial states under three noise channels was studied. It was showed that: $(i)$ The quantum decoherence,
for time $\lambda t\leq\lambda\widetilde{t}$, can also have the
contribution to the evolution of the system, which is contrary to
that in inertial frame, so we can get the conclusion that the
Unruh effect can induce that the quantum decoherence happens in this
time interval. $(ii)$ When $\lambda t\geq\lambda\widetilde{t}$, the
system's evolution also depends on the classical decoherence for the
bit flip and phase-bit flip, which is different from these three noise channels in inertial frames. But this conclusion doesn't apply to the Phase flip channel. $(iii)$ As the
Unruh temperature increases, the transition time  will be bigger for the
phase flip but be smaller for the bit flip compared with that in
inertial frame. However, this transition time does not depend on the
Unruh effect for the phase-bit flip. And $(iv)$  the Unruh effect can
also induce that the quantum discord becomes more than the classical
correlation in a little time interval near the
$\lambda\widetilde{t}$, the higher the Unruh temperature is, the longer
the interval is. Although the quantum and classical
decoherences, for some noise channels, may occurs when $\lambda
t\leq\lambda\widetilde{t} $ and $\lambda t\geq\lambda\widetilde{t}
$, respectively, the classical decoherence always dominates the
evolution of system when $\lambda t\leq\lambda\widetilde{t} $, while
the quantum decoherence makes primary contribution to system's
evolution when  $\lambda t\geq\lambda\widetilde{t} $.

\begin{acknowledgments}

This work was supported by the National Natural Science Foundation of China under Grant Nos. 11175065, 10935013; PCSIRT, No. IRT0964; the Hunan Provincial Natural Science Foundation of China under Grant No. 11JJ7001; the Construct Program of the National  Key Discipline; and the Project of Knowledge Innovation Program (PKIP) of Chinese Academy of Sciences, Grant No. KJCX2.YW.W10.

\end{acknowledgments}

\end{document}